**Saggu, A., Ante, L., & Demir, E. (2024). Anticipatory Gains and Event-Driven Losses in Blockchain-Based Fan Tokens: Evidence from the FIFA World Cup. Research in International Business and Finance, Volume 70, Part A, 102333.**

# Anticipatory Gains and Event-Driven Losses in Blockchain-Based Fan Tokens: Evidence from the FIFA World Cup

This Version: March 23, 2024

Aman Saggu [ab]*

[a] *Business Administration Division, Mahidol University International College*

[b] *Blockchain Research Lab*

Lennart Ante [b]

[b] *Blockchain Research Lab*

Ender Demir [c]

[c] *Department of Business Administration, Reykjavik University*

**Abstract:** National football teams increasingly issue tradeable blockchain-based fan tokens to strategically enhance fan engagement. This study investigates the impact of 2022 World Cup matches on the dynamic performance of each team's fan token. The event study uncovers fan token returns surged six months before the World Cup, driven by positive anticipation effects. However, intraday analysis reveals a reversal of fan token returns consistently declining and trading volumes rising as matches unfold. To explain findings, we uncover asymmetries whereby defeats in high-stake matches caused a plunge in fan token returns, compared to low-stake matches, intensifying in magnitude for knockout matches. Contrarily, victories enhance trading volumes, reflecting increased market activity without a corresponding positive effect on returns. We align findings with the classic market adage "buy the rumor, sell the news," unveiling cognitive biases and nuances in investor sentiment, cautioning the dichotomy of pre-event optimism and subsequent performance declines.

*Keywords:* Cryptocurrency; Blockchain; Fan Tokens; Football; Efficient Markets Hypothesis; Intraday Data.

*JEL Classification:* G12; G14; G41; L83.

*Corresponding author: Aman Saggu, Business Administration Division, Mahidol University International College, 999 Phutthamonthon Sai 4 Rd, Salaya, Phutthamonthon District, Nakhon Pathom 73170, Thailand, Telephone: 0066 27005000, Fax: 0066 24415091, Email address: asaggu26@gmail.com; aman.sag@mahidol.edu.

# 1   Introduction

The advent of blockchain-based fan tokens signifies a pioneering innovation at the intersection of finance, sports, and technology. Major organizations such as renowned sports clubs (e.g., FC Barcelona and Manchester City), well-known sports franchises (e.g., Formula 1 and the Davis Cup), and large corporations (e.g., Aston Martin and Alfa Romeo) have issued this novel digital asset – fan tokens – on public blockchains with a pre-determined supply, setting a fixed price for initial distribution to interested investors and stakeholders. These fungible utility tokens are tradeable on secondary markets and grant owners voting rights in specific organizational decisions, access to exclusive rewards, and other interactive opportunities, effectively deepening the influence, participation and engagement of fans with the organization (Ante et al., 2023; Demir et al., 2022).[1] By 2023, over 139 tradable fan tokens have been issued, having a combined market capitalization of more than USD 325 million (Rocketfan, 2023).

The FIFA 2022 World Cup (hereafter World Cup), an international football tournament, was held from November 20, 2022, to December 18, 2022 in Qatar. In the six months (or one month) leading up to this high-profile event, football fan tokens tied to the participating national teams (hereafter fan tokens) underwent remarkable surges in value.[2] Argentina's token (ARG) appreciated by a staggering 953% (or 94%), Portugal's token (POR) soared by an impressive 1,038% (or 84%), and Brazil's fan token (BFT) rose a notable 324% (or 7%).[3] These dynamics underscore the tangible impact of major sports events on the valuation of associated digital assets. This remarkable fan token growth defied the broader cryptocurrency bear market, substantiated by established Sentix Bitcoin and Alternative.me sentiment indicators (Saggu, 2022).[4] Interestingly, the World Cup coincided with extreme market turbulence following the crash of the FTX cryptocurrency exchange on November 11, 2022. Indeed, benchmark cryptocurrencies Bitcoin and Ethereum declined -19% and -5%, respectively, in one month preceding the World Cup – highlighting fan token resilience and growth amidst an otherwise declining cryptocurrency market.

Given the absence of national football team listings on stock markets, a considerable corpus of previous research has analyzed the ramifications of football match outcomes – specifically victories and defeats

---

[1] A utility token is a type of digital token issued on a blockchain that provides users with access to a product or service. Utility tokens are not designated as investments per se, but they can increase (decrease) in value based on supply and demand on secondary markets, which in turn may be tied to the success of the platform or service.
[2] In this paper, the term "football" is used to refer to the sport commonly known as soccer in the United States. By using the term "football," we adopt the European convention, where the sport is predominantly referred to as such.
[3] The token price changes are calculated using token price data from CoinGecko.com for the six (one) month period between June 20, 2022 and November 20, 2022 (October 20, 2022 and November 20, 2022). Only four national teams participating in the World Cup had associated fan tokens during the 2022 World Cup: Argentina (ARG), Portugal (POR), Brazil (BFT) and Spain (SNFT).
[4] The Sentix Bitcoin Sentiment (Alternative.me) index overwhelmingly indicated a bear market (indication of extreme fear or fear) over this period.

– on national stock market indices as representative proxies (cf. Ashton et al., 2011, 2003; Edmans et al., 2007; Kaplanski and Levy, 2010; Klein et al., 2009b, 2009a; Renneboog and Vanraband, 2000; Scholtens and Peenstra, 2009). This methodological approach is inherently limited due to the asynchrony of match times and trading hours of national stock markets (i.e., the stock market may not be trading during match times). As a result, they analyze the impact of the next available trading day. Several studies have assessed impacts on non-football stocks actively traded during matches (cf. Curatola et al., 2016; Ehrmann and Jansen, 2016). However, these studies are constrained by lack of direct association between selected stocks and national football teams. Further studies have delved into the impacts on the stock prices of publicly listed football teams – none of which are national football teams – yet they invariably face similar limitations regarding incongruities between match times and stock market trading hours (cf. Bernile and Lyandres, 2011; Demir and Danis, 2011; Geyer-Klingeberg et al., 2018; Palomino et al., 2009). A recent study explores the impact of football match outcomes on football fan tokens; however, the scope of research is limited to daily data (see Demir et al., 2022). Indeed, Bariviera and Merediz-Solà (2021) highlight that only 14% of cryptocurrency research papers employ intraday data. Our study contributes to this limited body of research by using minute-by-minute data.

In the context of our study on the dynamics of fan tokens amidst the 2022 FIFA World Cup, it is imperative to consider the broader financial mechanisms and market behaviors of the fan token market that intersect with the phenomena observed. The inception of the market for blockchain-based tokens can be traced back to initial coin offerings (ICOs), which emerged as a pivotal funding mechanism for crypto projects (Howell et al., 2020). These tokens sales often encountered significant underpricing and subsequent price surges after being listed on exchanges (Benedetti and Kostovetsky, 2021) and were exposed to ownership-user conflicts that could hinder investment (Cong et al., 2022). The phenomenon of token sales has continued to evolve, so that in the context of fan token sales we now refer to the phrase fan token offering (FTO).

The crypto asset market differentiates itself from traditional financial markets through its valuation dynamics. Unlike stocks, which are valued based on cash flows, crypto assets derive their value from their ability to facilitate transactions, user adoption, and network effects, making them appealing for speculative investments despite their volatility (Nadler and Duo, 2020; Cong et al., 2021). As the market evolved, it branched into various niches, each with its unique characteristics (Yousaf et al. 2023; 2024). Studies have highlighted the complex interplay between different types of tokens and traditional financial assets. For instance, while some football fan tokens show minimal correlation with football club stocks (Ersan et al., 2022; Scharnowski et al., 2023), lending and borrowing tokens demonstrate a stronger connection with bank stocks during uncertain times (Yousaf et al., 2022). This evolving landscape underscores the diverse impacts and relationships within the cryptocurrency sector, further complicated by market sentiment and news (Foglia et al., 2024).

Recent research by Assaf et al. (2024) on bubbles in fan token markets adds another layer to our understanding, indicating the prevalence of speculative bubbles that could influence market dynamics. Ante et al. (2024b) further highlights that fan tokens effectively engage fans, with an average of 4,003 participants per poll across 3,576 polls, demonstrating their utility in democratizing decision-making in sports and esports. This body of work collectively illustrates how the cryptocurrency market has matured from its initial phase of ICOs to a complex ecosystem with multiple niches, each influenced by distinct factors and exhibiting unique relationships with traditional financial markets.

In light of previous research, this paper investigates the efficiency of football fan tokens in responding to unfolding information concerning World Cup matches. The study analyzes fan tokens both directly associated with and issued by national football teams. This specificity addresses concerns in earlier studies which relied on proxies for national football teams, potentially skewing the nature of the relationship. Moreover, we leverage the continuous – operational 24/7 – nature of the cryptocurrency market, analyzing the performance of football fan tokens in real-time during matches using intraday data at the one-minute frequency, transcending the constraints of daily data aggregation in previous studies.[5] In pursuance of our objectives, we empirically investigate if: (1) victories (defeats) elicit positive (negative) effects on fan-token returns and trading volumes in real-time; (2) disparities in magnitude of returns and trading volume responses between victories and defeats; (3) divergences in magnitude of response can be explained by match importance (low-stake versus high-stake) and critical tournament phases (i.e., knockouts).

This paper uncovers insights into the behavioral dynamics of fan token markets during different phases of World Cup matches using three refined specifications of event-study models. Firstly, in the pre-match phase, we reveal a modest but statistically significant increase in cumulative abnormal returns (CARs), indicating market anticipation, and a rise in cumulative abnormal volumes (CAVs), reflecting heightened market activity, speculation and investor pre-positioning. Secondly, we demonstrate a decline in CARs during the half-time and second-half segments, associated with high trading volumes, suggesting market reactions to real-time match developments and a peak in investor response. Thirdly, we discern that the post-match segment experienced a further diminution in CARs and sustained significant trading volumes, indicating investor concentration on ultimate match outcomes which align with market phenomena such as post-event drift and the over-reaction hypothesis.

We subsequently investigate principal determinants influencing fan token CARs and CAVs during matches. We find that victories and defeats have observable effects on football fan token CARs and CAVs, with defeats having a more significant and larger magnitude impact. Our findings are consistent with asymmetries observed in existing literature on the impact on national stock markets (Ehrmann and

---

[5] Ehrmann and Jansen (2015) stands out as the only study deploying intraday data, albeit exclusively for a singular cross-listed stock. This specificity stemmed from the inherent limitations of stock trading continuity and the lack of direct ties to a football team in available options, restricting the scope of their analysis.

Jansen, 2016) and crypto assets (Demir et al., 2022). We identify that the impacts of match outcomes on fan token CARs are significantly heightened in high-stakes matches (round of 16, semi-final, final) compared to lower-stakes group stages, likely due to factors such as intensified emotional and temporal commitments from teams and fans, elevated media spotlight, and the enduring impacts on a team's legacy. We reveal that defeats in high-stakes matches, especially knockouts, are detrimental to fan token CARs, causing sharp declines in value, evidenced by a -50.74% to -59.43% additional decline using robust MM-estimator models. Further, we find nuanced insights into the impacts on fan token CAVs, revealing that victories have a robust, positive, and statistically significant influence, contrasting the more muted impacts of defeats. The insignificance of match stakes and knockout variables on CAVs indicates a dissociation between fan emotional and psychological commitment and trading activity, which allows trading volumes to sustain less impact even when CARs substantially depress in value during high-stakes defeats. Our investigation into fan token returns throughout World Cup matches unveil a significant pattern consistent with the "buy the rumor, sell the news" market maxim.

Lastly, we reveal a substantial surge in fan token returns in the six months preceding the World Cup, with a significant CAR of 211%, reflecting heightened investor anticipation. However, this optimistic momentum waned as the World Cup approached, with CARs diminishing but still positive, indicating a moderation in enthusiasm due to anticipated uncertainties surrounding match outcomes. Remarkably, the commencement of the World Cup marked a profound plunge in CARs, reaching -173% and further plummeting to -1,000% towards the tournament's end. This sharp transition in investor sentiment indicates risk-mitigation strategies, where investors capitalize on pre-event hype and liquidate positions as actual matches commence, illustrating complex fan token behavior during high-profile sporting events.

This study highlights the pivotal role of foundational behavioral finance principles, in the evolving context of fan tokens, emphasizing the significant influence of fan sentiment and emotional attachment on the market dynamics of these digital assets (Barberis and Thaler, 2003; Shleifer, 2000). The insights garnered reinforce the importance of recognizing the impact of sentiment-driven decisions and fan loyalty in fan token and similar unconventional markets. The evident association between the emotional fervor associated with football outcomes and consequential financial decision-making pertaining to related assets such as fan tokens supports the theory that deviations from rationality, fueled by emotional influences and cognitive biases, are especially prevalent in this domain. This contrasts with the efficient markets hypothesis, which proposes that markets are inherently efficient, reflecting all available information in asset prices (Fama, 1970), implying that publicly available football match outcomes should be rapidly integrated into asset prices. Meanwhile, the emerging domain of fan tokens, defined by their multifunctional and financialized nature coupled with increased trust and efficiency (Ante et al., 2023), serves as a fertile ground for further exploration into the intricate interrelation between sports and financial markets. By embracing behavioral considerations in the analysis of such assets, this study not only advances our understanding of unique market constructs but also opens avenues for further research into the intersectionality of sports sentiment and financial market dynamics.

## 2 Literature Review

### 2.1 Interplay between Football Match Outcomes and Financial Market Responses

Research examining the influence of football match results on financial markets has predominantly focused on their consequential effects on stock market dynamics. An early study by Renneboog and Vanraband (2000) found evidence that the share prices of football clubs publicly traded on the United Kingdom's (UK) major financial markets, namely the London Stock Exchange and the Alternative Investment Market, witnessed a significant 1% rise in abnormal returns on the first trading day following a match victory. Ashton et al. (2003), building on this foundation, demonstrated that the performance of the England national football team in the World Cup had significant repercussions on the nation's primary stock market index, the FTSE100, arguing that sports results impact investor psychology at a national level. Further broadening the geographical scope of the research, Edmans et al. (2007) found that this phenomenon extended beyond the UK, revealing asymmetries, as an elimination-stage defeat of a national team in the World Cup was associated with a significant negative abnormal return of -0.49% on the respective national stock market indices of 39 countries, while victories were statistically insignificant. The research supports the hypothesis that national sports outcomes significantly influence investor sentiment at a macro level. Nevertheless, the research landscape is not without its controversies.

Klein et al., (2009b, 2009a)challenged the validity of findings by Ashton et al. (2003) and Edmans et al. (2007), demonstrating that when controlling for surprise results based on probable outcome expectations from betting odds and when expanding the sample, the significant effects earlier observed were not replicable. In response to this critique, Ashton et al., (2011) conceded the impact of national team victories and defeats on national stock market indices had diminished over time. This concession introduced a novel consideration into the discourse – a time-varying element in the impact of sports outcomes on national stock markets. Accounting for these dynamics, Kaplanski and Levy (2010) argued that investors could not capitalize on the localized effect of a football victory due to the inherent uncertainty of the outcome. However, investors could leverage globally pervasive negative effects triggered by consecutive national-level defeats in international tournaments such as the World Cup. This intriguing hypothesis hints at the possibility of investors exploiting a systematic negative influence induced by aggregate sports outcomes on global stock market dynamics.

Research has also explored the phenomenon of asymmetric responses. A study by Palomino et al. (2009) highlighted an intriguing aspect of this, observing that non-national football clubs listed on the London Stock Exchange experienced abnormal returns of 0.88% (-1.01%) following victories (defeats) over a three-day span. Interestingly, the market responded more swiftly in one day (three days) to good (bad) news associated with victories (defeats). Scholtens and Peenstra (2009) also documented significant positive (negative) national stock market reactions to victories (defeats) and heightened reaction to defeats, European matches, and unforeseen results. Bernile and Lyandres (2011) ventured to demystify this asymmetry. They posited that investor sentiment towards European football clubs is subject to bias, attributable to inflated ex-ante expectations when unmet culminate in acute disappointment. This emotional letdown, they proposed, precipitates negative post-match abnormal returns, adding a layer of psychological nuance to the understanding of market responses to sports outcomes. In a dichotomy between local and international matches, Demir and Danis (2011) identified that victories in local football matches exerted a more pronounced influence on the share prices of Turkish football clubs than victories at European Cup matches. The tendency towards localized market reaction is potentially attributable to the closer emotional connection of domestic investors with local matches.

The initial studies probing the impact of football match outcomes on financial markets paved the way for further explorations, yielding intriguing insights. Curatola et al. (2016) unveiled the sector-specific effect of World Cup matches, demonstrating that matches exhibited significant influence exclusively on highly liquid financial sector stocks in the USA, with insignificant responses for other sectors. Building on this discourse, Ehrmann & Jansen (2016) analyzed the peculiar underpricing pattern of STMicroelectronics shares listed on Paris and Milan exchanges during two football matches featuring France and Italy. This distinct anomaly, identified as "mood effects," emerged alongside an escalating probability of the competing teams facing defeat during the matches. Extending upon these sector-specific and company-specific investigations, Geyer-Klingeberg et al. (2018) conducted a meta-analysis encompassing 37 studies, validating the insignificant impact of victories on stock returns in preceding studies. However, after accounting for publication bias, they revealed more muted negative effects of national team (individual club) defeats with an impact of -0.05% (-0.39%) on national stock markets.

Studies have also delved into the intricate interplay between investor psychology and football matches. An early study (Wann et al., 1994) revealed that fans who exhibit high identification with a team believe they can influence team performance. This intense fan engagement induced emotional responses to match outcomes, positive for victories and negative for defeats. Notably, although past team success contributed to higher fan identification, individual match outcomes did not impact identification levels. A later study by Yu and Wang (2015) dissected the sentiments in U.S. football fans' tweets during the 2014 World Cup, discerning the alignment of reactions with the U.S. team's performance. Negative emotions surged following the concession of goals, and successful scoring sparked joy and anticipation. Interestingly, matches not involving the United States evoked positive sentiments in the tweets, corroborating the disposition theory of fan response and team affiliation.

This exploration of sentiment-driven responses to match outcomes extended by Demir and Rigoni (2017) scrutinized the schadenfreude effect–satisfaction derived from the misfortune of other teams– on the stock prices of the Italian football clubs Roma and Lazio. They observed investor reactions oscillate positively with their preferred team's victories and negatively with defeats. However, the rival team's performance influenced stock prices only when they coincided with a favorite team's defeat. An unexpected victory or defeat by the rival team either neutralized or amplified, respectively, the negative market reaction to the favorite team's defeat, implying complex interdependence of investor sentiments and rational expectations in shaping the stock performance of listed football clubs. More recently, Truong et al. (2021) reported a 0.28% decline in Vietnamese stock returns following an international football defeat. This trend intensified for small-cap stocks over a period of heightened fan engagement from 2015 to 2020. This evidence underscores the pivotal role of emotional biases in the determination of market performance, highlighting the profound influence of football match outcomes on investor psychology and, subsequently, on financial markets

## 2.2 Interplay between Football Match Outcomes and Football Fan Tokens

Fan tokens represent a pioneering deployment of blockchain technology within the sporting industry, particularly football. Fan tokens grant holders participatory governance rights in certain decisions, exclusive rewards, and interactive opportunities within the organization. This structure amplifies holders' influence, promotes active participation, and heightens engagement. Essentially, fan tokens intensify the bond between fans and their favored teams by offering exclusive perks, decision-making authority, and opportunities for interaction (Ante et al., 2023; Demir et al., 2022). FC Barcelona's BAR token issuance in June 2020 initiated the introduction of this novel asset class: fan tokens. It attracted institutional and retail investors, major football clubs, and enthusiast fans alike (Ante et al., 2023). As of June 2023, the fan token marketplace had over 139 tradable tokens, amassing a cumulative market capitalization of USD 325 million (Rocketfan, 2023). These digital assets permeate diverse sporting sectors, including but not limited to football clubs, eSports organizations, basketball clubs, and motorsport teams. Notwithstanding the rapid escalation in their popularity, academic discourse concerning fan tokens is still in the early stages, a direct consequence of the novelty and innovation intrinsic to this emergent class of digital assets.

A pioneering study by Demir et al. (2022) revealed that victories (defeats) in the UEFA Champions League were associated with positive (negative) abnormal fan token returns, with the impact of defeats nearly twice as pronounced – aligning with asymmetries observed in national stock market behavior. Notably, this phenomenon was absent in domestic or Europa League matches, indicating the heightened influence of the Champions League's prestige on fan sentiment. Investigating fan token risk and return characteristics, Mazur & Vega (2022) revealed impressive initial trading day returns of approximately 150%. However, these tokens underperformed in the longer term compared to benchmarks like Bitcoin,

DeFi, NFT, and Meme tokens despite high trading volumes on platforms like Binance. Intriguingly, fan tokens outperformed shares of publicly traded European football clubs. Building on this research, Scharnowski et al. (2022) constructed an equally weighted index of football fan tokens. Their findings indicated that these fan tokens shared similarities with volatile cryptocurrencies than traditional assets like publicly listed football club stocks. They also noticed an evolving market where investor attention significantly impacted returns, and unexpected match defeats began influencing fan token prices. They concluded that most fan token prices exhibited weak-form efficiency, with minor inefficiencies for smaller clubs.

Adding a different perspective, Ersan et al. (2022) investigated the dynamic interconnectedness between football fan tokens and corresponding football club stocks. They found that fan tokens transmitted shocks to other tokens and stocks. However, stocks generally acted as receivers of these shocks. Over time, however, the total connectedness index, which measures interactions among assets, significantly decreased, indicating a decline in interconnectedness. Expanding the analysis to football and additional sports fan tokens, Vidal-Tomás (2023) found that investment in Socios tokens yielded superior returns compared to targeted investments in specific team tokens, further documenting the end of the fan token bubble in late 2021.

Additionally, Lopez-Gonzalez & Griffiths (2023) revealed the gambling-like features of BAR tokens and their potential to exploit fans' loyalty. They argued fan tokens encourage constant engagement, turning fandom into a competitive and potentially exploitative activity. More recently, Ante et al. (2024a) observed a 0.8% decline in fan token returns during football matches. Concurrently, Manoli et al. (2024) employed a survey methodology to investigate the motivations of fan token investors, identifying two distinct cohorts: one views fan tokens as symbolic of their allegiance to football clubs, while the other perceives them purely as investment vehicles. This bifurcation highlights the dual nature of fan tokens as both emotional and financial assets within the sports ecosystem.

# 3 Data

## 3.1 Cryptocurrency returns

Our empirical dataset encompasses the 2022 FIFA World Cup, from 00:00 on November 20, 2022, to 23:59 on December 18, 2022 (UTC). Intraday price data at the one-minute frequency is sourced from the Gate.io cryptocurrency exchange for four fan tokens associated with national teams competing in the event: Argentine Football Association (ARG); Brazil National Football Team (BFT); Portugal

National Team (POR) and Spain National (SNFT).[6][7][8] Our dataset is further supplemented with intraday price data for Bitcoin (BTC) and Chiliz (CHZ), sourced from Gate.io, serving as benchmark markets for our analysis.[9] All price returns are calculated as the first difference of the natural logarithm of the one-minute closing price relative to the previous minute, i.e., $r_t = 100 * \ln(p_n/p_{n-1})$.

## 3.2 Performance metrics

The dataset is enriched with performance metrics for each match involving the four national teams, utilizing data from transfermarkt.com and fifa.com. These metrics include the match stage (e.g., group stage, round of 16, quarterfinals, semi-finals, and final), the number of goals scored, and the total playtime (incl. overtime, extra time, and penalties), with all times converted to Universal Coordinated Time (UTC) for uniformity. Betting odds from oddsportal.com are also integrated into our dataset, measuring the extent to which match outcomes were expected or surprising. Given the high rankings of Argentina, Brazil, Portugal, and Spain (3rd, 1st, 9th, and 7th, respectively, among the 211 nations in the FIFA men's world ranking as of October 6, 2022), these teams were predominantly perceived as favorites in their respective matches (FIFA, 2022).

In line with the methodologies of Palomino et al. (2009) and Scharnowski et al. (2022), we determine ex-ante victories (hereafter victories) and ex-ante defeats (hereafter defeats) by making appropriate adjustments for bookmaker fees, commonly referred to as 'over-roundness' and classifying expected victories and defeats if differences between victory and defeat probabilities are greater than 30%.[10] An overview of the 21 matches under consideration in our study, accompanied by their respective results, is presented in Table 1.

---

[6] The Peruvian National Football Team (FPFT) and Italian National Football Team (ITA) fan tokens are not applicable to this study as the corresponding teams from Italy and Peru did not qualify for the tournament.

[7] The selection of Gate.io as the primary source of cryptocurrency price data is motivated by its comprehensive provision of freely accessible price data at an intraday one-minute frequency for all four fan tokens as well as the reference markets, covering the entire duration of the World Cup. This selection facilitates replicability of our study using publicly accessible data. At the onset of the World Cup on December 18, 2022, Gate.io held the seventh position among global cryptocurrency exchanges in terms of trading volume, liquidity, weekly visits, number of markets, and the diversity of coins traded, as per the rankings published by coinmarketcap.com. The trading volumes on Gate.io exceeded those observed on other prominent cryptocurrency exchanges such as Bitstamp, KuCoin, Bitfinex, and Binance US.

[8] In line with Alexander and Dakos (2020), who underscore the importance of utilizing traded data from cryptocurrency exchanges, our study employs directly traded prices from the Gate.io exchange. Advancing this discourse, Vidal-Tomás (2022) points out that the selection of an exchange can influence the statistical characteristics of cryptocurrency price data. While Gate.io is not amongst the largest cryptocurrency exchanges in the world, it consistently ranked within the top 8 to 14 exchanges globally by trading volume over our sample period, according to Coinmarketcap and CoinGecko. Gate.io's notable trading activity helps to alleviate potential concerns regarding disparities in research outcomes that might arise from the choice of data source.

[9] Chiliz (CHZ) operates as a cryptocurrency and blockchain technology affiliated with the Socios.com platform which seeks to enhance fan engagement and interaction. The Socios.com platform serves as the primary ecosystem allowing for the procurement of fan tokens via the use of CHZ. Once purchased, these tokens are subsequently stored in a Socios.com-designated digital wallet. One of the key attributes of these fan tokens is their fungibility, which allows them to be traded on diverse cryptocurrency exchange platforms. Chiliz emerges as an optimal reference market for the examination of football fan tokens due to its status as the native cryptocurrency of the Socios.com fan engagement platform. This relationship implies that the majority of fan tokens are procured utilizing CHZ, thus closely aligning the price of CHZ with the demand for fan tokens. Furthermore, Chiliz's stature as a preferred cryptocurrency for football fan tokens, the considerable liquidity of its pool, and its established standing with a demonstrable track record, further enhance its role as a suitable reference market in the analysis of football fan tokens.

**Table 1: Descriptive statistics for 2022 World Cup matches for teams with cryptocurrency fan tokens**

| ID | Date | Time | Match | Match Stage | Outcome | Score |
|---|---|---|---|---|---|---|
| 1 | Nov 22, 2022 | 10:00 | Argentina vs. Saudi Arabia | Group stage 1 | Defeat$^S$ | 1:2 |
| 2 | Nov 23, 2022 | 16:00 | Spain vs. Costa Rica | Group stage 1 | Victory | 7:0 |
| 3 | Nov 24, 2022 | 19:00 | Brazil vs. Serbia | Group stage 1 | Victory | 2:0 |
| 4 | Nov 24, 2022 | 16:00 | Portugal vs. Ghana | Group stage 1 | Victory | 3:2 |
| 5 | Nov 26, 2022 | 19:00 | Argentina vs. Mexico | Group stage 2 | Victory | 2:0 |
| 6 | Nov 27, 2022 | 19:00 | Spain vs. Germany | Group stage 2 | Draw | 1:1 |
| 7 | Nov 28, 2022 | 16:00 | Brazil vs. Switzerland | Group stage 2 | Victory | 1:0 |
| 8 | Nov 28, 2022 | 19:00 | Portugal vs. Uruguay | Group stage 2 | Victory | 2:0 |
| 9 | Nov 30, 2022 | 19:00 | Argentina vs. Poland | Group stage 3 | Victory | 2:0 |
| 10 | Dec 1, 2022 | 19:00 | Spain vs. Japan | Group stage 3 | Defeat$^S$ | 1:2 |
| 11 | Dec 2, 2022 | 19:00 | Brazil vs. Cameroon | Group stage 3 | Defeat$^S$ | 0:1 |
| 12 | Dec 2, 2022 | 15:00 | Portugal vs. South Korea | Group stage 3 | Defeat$^S$ | 1:2 |
| 13 | Dec 3, 2022 | 19:00 | Argentina vs. Australia | Round of 16 | Victory | 2:1 |
| 14 | Dec 5, 2022 | 19:00 | Brazil vs. South Korea | Round of 16 | Victory | 4:1 |
| 15 | Dec 6, 2022 | 15:00 | Spain vs. Morocco | Round of 16 | Defeat$^S$ | 0:0 (0:3) |
| 16 | Dec 6, 2022 | 19:00 | Portugal vs. Switzerland | Round of 16 | Victory | 6:1 |
| 17 | Dec 9, 2022 | 15:00 | Brazil vs. Croatia | Quarter-finals | Defeat$^S$ | 1:1 (4:2) |
| 18 | Dec 9, 2022 | 19:00 | Argentina vs. Netherlands | Quarter-finals | Victory | 2:2 (4:3) |
| 19 | Dec 10, 2022 | 15:00 | Portugal vs. Morocco | Quarter finals | Loss$^S$ | 0:1 |
| 20 | Dec 13, 2022 | 19:00 | Argentina vs. Croatia | Semi-finals | Victory | 3:0 |
| 21 | Dec 18, 2022 | 15:00 | Argentina vs. France | Final | Victory | 3:3 (4:2) |

Note: All dates and times conform to Universal Coordinated Time (UTC). For clarity, the sequence of teams in each match has been adjusted to place the team linked to a fan token at the forefront. Outcomes deemed surprising based on betting odds are annotated with a superscript 'S'. These outcomes (including surprises) are considered from the perspective of the four-fan token-associated national teams (Argentina, Brazil, Portugal, and Spain). Scores enclosed in parentheses denote final outcomes determined by penalties.

## 4 Empirical Results

### 4.1 Preliminary analysis

Figure 1 presents a salient trend of declining returns for all four fan tokens over the duration of the World Cup. The BFT token experienced the most pronounced decline (-95%). Conversely, the ARG token experienced the least severe decline with a -52% return, despite the associated team Argentina winning the World Cup. The ARG token briefly appreciated closer to its initial value (-6%) during the semi-final phase, only to experience a subsequent decline. This data prompts further rigorous exploration into the determinants influencing fan token returns during football matches in the World Cup, focusing on factors like the associated team's performance and broader market dynamics. Our initial findings align with Ehrmann & Jansen's (2016) hypothesis that successive defeats across a tournament may trigger a negative mood effect in the broader stock market, which our preliminary estimates suggest may apply to cryptocurrency fan tokens directly associated with national football teams.

---

[10] For example, the match between Argentina and Saudi Arabia had the following odds: *victory Argentina* = 1.12, *draw* = 9.21 and *victory Saudi Arabia* = 25.52. Converted into percentages, this corresponds to 89.3%, 10.9% and 3.9%, which in total equals 104.1% and includes the fee of the betting provider (4.1%). Adjusted for the fee, this leads to odds of 85.8% (victory), 10.4% (draw) and 3.8% (defeat) for Argentina. As the difference between victory and defeat is higher than 30%, the match is classified as an ex-ante expected victory. Therefore, the victory of Saudi Arabia represents an "unexpected defeat" for Argentina.

**Figure 1: Cumulative fan token returns during the 2022 World Cup**

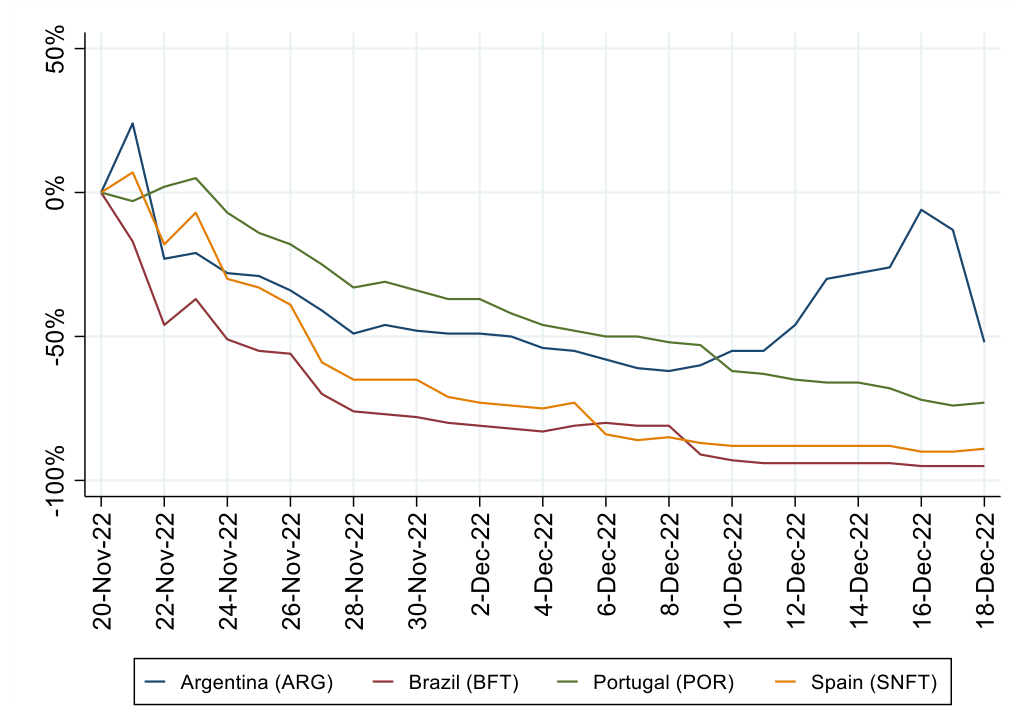

### 4.2 Methodology

Table 2 presents the quantified effects of World Cup matches on the cumulative average abnormal returns and trading volumes of associated fan tokens. The analytical framework consists of three models: (a) constant mean return model, (b) market model with Bitcoin (BTC) as the benchmark market, and (c) market model with Chiliz (CHZ) as the sector-specific benchmark market. The first objective of this methodological approach is to empirically assess the significance of market effects across event windows with respect to the anticipated market performance, as outlined by (Boehmer et al. (1991). The second objective is to gauge the robustness of our model non-contingent on its specifications, thereby fortifying the integrity and veracity of our estimates with three alternative models (MacKinlay, 1997). By implementing these three alternate model specifications, we ensure our findings are not merely an artifact of a particular model, but provide a comprehensive and dependable analysis of the impact of World Cup matches on fan token dynamics.

In the market model, the logged fan token return $R_{i,t}$, for fan token $i$ at time $t$ is calculated using Equation (1):

$$R_{i,t} = \alpha_i + \beta_i R_{m,t} + \varepsilon_{i,t}, \quad (1)$$

where $\alpha_i$ is the intercept term specific to fan token $i$ reflecting the observed token log return, $R_{m,t}$ represents the log return of the reference market $m$ (either BTC or CHZ) at time $t$, $\beta_i$ estimates the sensitivity coefficient of fan token $i$ to the reference market's returns, and $\varepsilon_{i,t}$ is the error term. The abnormal return $AR_{i,t}$ (Equation 2) for fan token $i$ at time $t$ is then defined as the difference between the actual return $R_{i,t}$ and the expected return from Equation (1).

$$AR_{i,t} = R_{i,t} - (\alpha_i + \beta_i R_{m,t}) \quad (2)$$

The cumulative abnormal return $CAR_{i,t}$ for fan token $i$ at time $t$ is calculated by summing up abnormal returns from time $t$ to $n$, and the cumulative average abnormal return $CAAR_{i,t}$ in Equation (3) is obtained by taking the average of $CAR_{i,t}$ from time $t$ to $n$. Excluding the reference market's coefficient, the market model transforms into the constant mean return model. Consistent with Ajinkya and Jain (1989) we employed a log (x+c) transformation to accommodate zero-inflated volume observations, adjusting observed trading volume (x) with a constant (c) of 1.

$$CAAR_{i,t} = \left(\frac{1}{N}\right) \cdot \sum_{t=1}^{n} CAR_{i,t} \quad (3)$$

The estimation window spans from -1,500 to -61 minutes for each match. We strategically select a duration of 1,440 minutes, equivalent to one day, for the estimation period. This duration serves three critical functions. Firstly, it affords us the opportunity to examine potential pre-match market reactions. Secondly, it mitigates any risk of event overlap, which could induce issues associated with cross-sectional correlations (MacKinlay, 1997). Finally, the selected period length ensures that the sensitivity of our results remains robust, uninfluenced by the period's duration (Armitage, 1995). Thus, our choice of estimation period aims to balance the necessity for comprehensive data capture with the requirement for methodological rigor. In line with Demir et al. (2022), we utilize both Bitcoin (BTC) and Chiliz (CHZ) as our reference markets.[11]

---

[11] Bitcoin dominates more than 50% of the cryptocurrency market in terms of market capitalization, substantiating its significance in the digital asset industry. On the other hand, CHZ acts as the proprietary token for the Socios platform and Chiliz blockchain, which holds a prominent position as the principal platform and infrastructure for fan tokens, as confirmed by research conducted by (Demir et al., 2022).

## 4.3 Influence of World Cup matches on fan token activity

Table 2 identifies distinct inter-match segments referred to as event windows, utilized for examining the repercussions of World Cup matches on the financial dynamics of fan tokens. The inter-match segments are categorized as follows: (i) Pre-Match: the period leading up to the match's kick-off, from $-60^{th}$ to $0^{th}$ minute. (ii) First Half: the initial segment of the match, approximately $0^{th}$ to the $45^{th}$ minute. (iii) Half-Time: the intermission between the segments, approximately $45^{th}$ to the $60^{th}$ minute. (iv) Second Half: the latter segment of the match, approximately $60^{th}$ to the $105^{th}$ minute. (v) Regular Match: the entirety of a standard match, excluding extensions like overtime, extra time, or penalties, approximately $0^{th}$ to $90^{th}$ minute. (vi) Full Match: this entire match duration, including extensions like overtime, extra time, and penalties, from $0^{th}$ to $\lambda^{th}$ minute, as $\lambda$ varies based on extensions. (vii) Post-Match: From the moment the match ends to 60 minutes post-match, from $\lambda^{th}$ minute to $\lambda^{th}+60$ minutes.

It is crucial to emphasize that the temporal demarcations provided are approximations designed to facilitate a coherent interpretation of the findings. While we use standardized time ranges for clarity of presentation and interpretability, the actual duration of each match segment can vary due to extensions or other factors (e.g., the first half may span from the $0^{th}$ to the $50^{th}$ minute for one match and $0^{th}$ to $53^{rd}$ minute for another). Accordingly, we define precise calculations for the commencement and conclusion of each segment of each individual football match and report within the framework of this standardized categorization (e.g., $0^{th}$ to $45^{th}$ minute) to ensure consistent data interpretation.[12]

---

[12] Additional minutes observed during the tournament are a consequence of the FIFA Referees Committee's calculation of the additional time to be allocated (FIFA, 2022).

**Table 2. Quantitative Effects of World Cup Matches on Fan Token Returns and Trading Volumes**

| Period | Returns | | | | | Trading volume | | | | |
|---|---|---|---|---|---|---|---|---|---|---|
| | CARs | SE | t-test | z-test | Pos. | CAVs | SE | t-test | z-test | Pos. |
| **(a) Constant Mean Return** | | | | | | | | | | |
| (i) Pre-match (-60 to 0) | 0.055% | 0.651% | 0.08 | 0.30 | 52% | 37.179 | 10.631 | 3.50*** | 3.04*** | 86% |
| (ii) First half (0 to 45) | -0.168% | 1.536% | -0.11 | -0.40 | 48% | 52.723 | 12.355 | 4.27*** | 3.49*** | 86% |
| (iii) Half time (45 to 60) | -1.631% | 0.863% | -1.89* | -2.66*** | 24% | 21.857 | 4.915 | 4.45*** | 3.56*** | 81% |
| (iv) Second half (60 to 105) | -7.952% | 2.598% | -3.06*** | -3.35*** | 14% | 91.960 | 14.543 | 6.32*** | 3.91*** | 90% |
| (v) Regular match (0 to 105) | -8.204% | 2.917% | -2.81*** | -2.69*** | 24% | 164.931 | 26.431 | 6.21*** | 3.95*** | 95% |
| (vi) Full match (0 to $\lambda$) | -10.483% | 4.064% | -2.58** | -2.42** | 29% | 193.226 | 32.547 | 5.94*** | 3.95*** | 95% |
| (vii) Post match ($\lambda$ to $\lambda$+60) | -4.363% | 3.144% | -1.39 | -1.03 | 38% | 63.449 | 16.336 | 3.88*** | 3.35*** | 90% |
| **(b) Market Model (Bitcoin)** | | | | | | | | | | |
| (i) Pre-match (-60 to 0) | 0.089% | 0.652% | 0.14 | 0.37 | 52% | 34.279 | 10.757 | 3.19*** | 2.76*** | 71% |
| (ii) First half (0 to 45) | -0.162% | 1.554% | -0.10 | -0.43 | 48% | 51.008 | 12.741 | 4.00*** | 3.35*** | 81% |
| (iii) Half time (45 to 60) | -1.656% | 0.868% | -1.91* | -2.69*** | 24% | 21.523 | 4.993 | 4.31*** | 3.53*** | 81% |
| (iv) Second half (60 to 105) | -7.924% | 2.596% | -3.05*** | -3.35*** | 14% | 93.072 | 14.901 | 6.25*** | 3.91*** | 90% |
| (v) Regular match (0 to 105) | -8.187% | 2.931% | -2.79** | -2.69*** | 24% | 163.148 | 27.443 | 5.94*** | 3.91*** | 90% |
| (vi) Full match (0 to $\lambda$) | -10.456% | 4.071% | -2.57** | -2.42** | 40% | 194.801 | 34.708 | 5.61*** | 3.91*** | 95% |
| (vii) Post match ($\lambda$ to $\lambda$+60) | -4.379% | 3.136% | -1.40 | -1.03 | 38% | 66.030 | 16.734 | 3.95*** | 3.39*** | 90% |
| **(c) Market Model (Chiliz)** | | | | | | | | | | |
| (i) Pre-match (-60 to 0) | 0.052% | 0.542% | 0.08 | 0.30 | 52% | 36.128 | 10.545 | 3.43*** | 2.94*** | 81% |
| (ii) First half (0 to 45) | -0.187% | 1552% | -0.12 | -0.43 | 48% | 51.369 | 12.217 | 4.20*** | 3.42*** | 86% |
| (iii) Half time (45 to 60) | -1.644% | 0.867% | -1.90* | -2.66*** | 24% | 21.355 | 4.927 | 4.33*** | 3.53*** | 81% |
| (iv) Second half (60 to 105) | -7.928% | 2.598% | -3.05*** | -3.35*** | 14% | 90.971 | 14.534 | 6.25*** | 3.91*** | 90% |
| (v) Regular match (0 to 105) | -8.211% | 2.926% | -2.81** | -2.66*** | 24% | 161.089 | 26.266 | 6.13*** | 3.95*** | 95% |
| (vi) Full match (0 to $\lambda$) | -10.485% | 4.071% | -2.58** | -2.42** | 29% | 190.272 | 32.286 | 5.89*** | 3.95*** | 95% |
| (vii) Post match ($\lambda$ to $\lambda$+60) | -4.395% | 3.142% | -1.40 | -0.99 | 43% | 62.638 | 16.289 | 3.85*** | 3.35*** | 90% |

Note: Table 2 presents a comprehensive analysis of the quantitative effects of World Cup matches on fan token Cumulative Abnormal Returns (CARs) and Cumulative Abnormal Volumes (CAVs) across various time periods surrounding the matches. The table is divided into three sections: (a) Constant Mean Return Model, (b) Market Model adjusted for Bitcoin, and (c) Market Model adjusted for Chiliz. For each section, we report outcomes for seven distinct periods: (i) pre-match, (ii) first half, (iii) half-time, (iv) second half, during the (v) regular match, for the (vi) full match duration, and (vii) post-match. Key metrics include CARs and their Standard Errors (SE), t-tests and Wilcoxon (1945) z-tests sign rank test for statistical significance, and the proportion of positive values (Pos.). The same structure is maintained for CAVs. Statistical significance is denoted using asterisks, where * $p < 0.10$, ** $p < 0.05$, and *** $p < 0.01$.

Table 2 presents estimates of Cumulative Abnormal Returns (CARs) and Cumulative Abnormal Volumes (CAVs) using the three alternative models outlined in Section 4.2: (a) the Constant Mean return model as a baseline for market neutrality; (b) the Market Model benchmarked against Bitcoin, to gauge relative importance against a major cryptocurrency; and (c) the Market Model benchmarked against Chiliz for a sector-specific gauge for more nuanced interpretations of fan token behavior

In the (i) pre-match phase, fan token CARs exhibited a slight but statistically significant increase of around 0.1% across all three models (models a to c). This subdued performance indicates restrained market anticipation or pre-positioning, implying trades had been predominantly positioned beforehand. Simultaneously, an elevated and statistically significant rise in CAVs signals heightened market activity and speculation, reflecting investor readiness for potential fluctuations in fan token prices triggered by each impending match.

Throughout the (ii) first half, CARs experienced a slight reduction of approximately -0.2% across all models (a to c), which, being statistically insignificant, denotes investors made mild adjustments to their positions, responding to initial match developments. In the (iii) half-time segment, CARs sharply and significantly declined between -1.6% and -1.7%, a steadfast pattern across different model specifications (a to c). The downward trend paralleled elevated, statistically significant trading volumes, signifying sustained investor engagement.

As each match progressed into the (iv) second half, CARs further precipitated, registering statistically significant drops of -7.9% to -8.0% across models (a to c). This decrease was coupled with a peak in trading volumes amongst the four inter-match segments (i to v), insinuating potent investor responses to real-time match developments during the final half of each match. This period represented a pinnacle of risk for football fan token returns, reflecting market reactions stemming from match developments and corresponding shifts in investor sentiment and expectations.

In the (vii) post-match segment, CARs experienced a further diminution, approximately -4.4%, although they ceased to maintain statistical significance. Concurrently, the persistence of significant trading volumes indicates investor attention concentrated on ultimate match outcomes, aligning with market phenomena such as the post-event drift (Ante, 2019) and the over-reaction hypothesis.

Overall, the (vi) regular match, excluding overtime and penalties, showcased significant CARs of around -8.2% and exhibited exceedingly high and significant trading volumes. When factoring in the full match duration, including overtime and penalties, negative returns intensified to around -10.5%. The events during this phase reflect broader market behavioral patterns and investors' reactions to the final outcomes of the matches.

The findings in Table 2 may appear counterintuitive at first glance. Therefore, we advance a series of interconnected hypotheses to explain the observed declines in fan token returns over inter-match segments. First, the six-month period preceding the World Cup experienced an extraordinary surge in fan token returns (see Section 1), continuing to edge up marginally in the (i) pre-match phase. Intriguingly, prices began declining from the start to the conclusion of each match. This trend aligns with the established market adage "buy the rumor, sell the news." Anticipating heightened attention to the World Cup, investors likely foresaw a concurrent rise in fan token prices. The liquidation of positions by investors at the start of each match mitigates risks related to backing a potentially losing team. This phenomenon is explored further in Section 5.

Secondly, cognitive biases, such as loss aversion and recency bias, can magnify market responses to negative news compared to positive news (Meng & Fu, 2020). Investors responding more vehemently to disappointing match developments may initiate more aggressive fan token sell-offs compared to acquisitions following positive match advancements.

Thirdly, drawing on the assertions of Kaplanski and Levy (2010), the singular nature of victory in football matches – i.e., only one team can win the World Cup – engenders cumulative negative sentiment from match stage defeats by all other participating teams. In the context of football fan tokens, as each team approaches defeat in a match, the corresponding fan tokens incur declines, fostering a sector-wide downturn in all fan token values.

Fourthly, Table 1 reveals that all matches labelled as defeats in the World Cup defied expectations set by betting odds (i.e., surprise losses). This disparity between expectation and reality may intensify the downward pressure on fan tokens associated with the losing team.

Fifthly, akin to the positive feedback loops articulated in game theory, a team gaining early advantages is more prone to maintaining that lead by running down the clock, triggering premature declines in fan tokens, which intensify as matches progress for the team falling behind. Lastly, analogous to the concept of search trees in computer science, the spectrum of potential outcomes condenses as each match unfolds, enhancing the predictability of the final result as time elapses.

## 4.4 Analyzing determinants

To deepen our investigation, we investigate the principal determinants influencing fan token CARs and CAVs throughout the full matches, as outlined in Tables 3 and 4, respectively. The foundational model (a) in Table 3 integrates dummy variables to account for match outcomes, specifically victories ($Win_D$) and defeats ($Loss_D$), aligning with Equation (4). The models refrain from distinguishing between expected and unexpected defeats, as, according to the betting odds delineated in Table 1, all defeats in the sample were unanticipated.

$$CAR_{i,t} = \alpha_1 + \beta_1(Win_D) + \beta_2(Loss_D) + \varepsilon_t \quad (4)$$

In terms of market reaction, estimates of Equation 4 using OLS in model (a) reveals that while victories ($\beta_1$) yield positive effects on corresponding football fan tokens, defeats ($\beta_2$) induce negative effects. Nevertheless, the magnitude of response to defeats is greater than that of victories. This finding aligns with existing literature on the impact on national stock markets (see Section 2.1) and broader cryptocurrency assets (see Section 2.2). However, the influences of victories and defeats does not attain statistical significance. The asymmetry in CAR results—significant for unexpected defeats and insignificant for largely anticipated victories—is inherently a feature of our dataset as outlined in Table 1. While presenting a minor limitation, it also advances the literature by revealing a divergence in investor behavior between expected and unexpected outcomes.

To reinforce the reliability of our findings, we conduct two robustness checks. Initially, model (f) utilizes a robust MM-estimator model, producing results that closely mirror our baseline model (a). Subsequently, by extending our exploration to Equation (5), we incorporate control dummies ($\sum_n \delta_n Z_{i,t}$) for match stages—specifically, the group stage, round of 16, quarter-finals, and finals—in Models (b) and (g). The outcomes persist in displaying statistically insignificant market reactions to both victories and defeats.[13]

---

[13] We excluded draws, the SNFT fan token, and semi-final matches from the respective models, serving as reference groups. This methodological choice is consistent with established literature in the fields of both stock performance and fan token dynamics (Edmans et al., 2007; Palomino et al., 2009) as well as fan tokens specifically (Demir et al., 2022).

**Table 3. Determinants of Cumulative Abnormal Returns During Full Match Period, OLS and Robust MM-Estimator Models**

|  | (a) Coef. (SE) | (b) Coef. (SE) | (c) Coef. (SE) | (d) Coef. (SE) | (e) Coef. (SE) | (f) Coef. (SE) | (g) Coef. (SE) | (h) Coef. (SE) | (i) Coef. (SE) | (j) Coef. (SE) |
|---|---|---|---|---|---|---|---|---|---|---|
| **Outcome** | | | | | | | | | | |
| (i) Victory | 0.0035 (0.0258) | 0.0199 (0.0562) |  | 0.0035 (0.1713) | -0.0277 (0.0235) | 0.0044 (0.1889) | -0.0177 (0.0775) |  | 0.0080 (0.1180) | -0.0257 (0.0825) |
| (ii) Victory (Low-Stake) |  |  | -0.1187 (0.1331) |  |  |  |  | -0.1176 (0.0881) |  |  |
| (iii) Victory (High-Stake) |  |  | -0.0796 (0.1389) |  |  |  |  | -0.0576 (0.0919) |  |  |
| (iv) Defeat | -0.1721 (0.1203) | -0.1577 (0.1350) |  | -0.0744 (0.1846) | -0.0744 (0.1220) | -0.1432 (0.1946) | 0.0103 (0.0799) |  | 0.0214 (0.1272) | 0.0227 (0.0863) |
| (v) Defeat (Low-Stake) |  |  | -0.1654 (0.1438) |  |  |  |  | -0.0772 (0.0952) |  |  |
| (vi) Defeat (High-Stake) |  |  | -0.3935** (0.1515) |  |  |  |  | -0.5770*** (0.1003) |  |  |
| (vii) Defeat & Knockout |  |  |  | -0.2281* (0.1261) | -0.3665* (0.1759) |  |  |  | -0.5074*** (0.0869) | -0.5943*** (0.0809) |
| $R^2$ | 0.2061 | 0.3229 | 0.3665 | 0.3342 | 0.4985 | 0.1082 | 0.3410 | 0.2443 | 0.1688 | 0.4510 |
| Adj. $R^2$ [pseudo $R^2$] | 0.1179 | 0.0327 | 0.2081 | 0.2167 | 0.2285 | [0.1692] | [0.7225] | [0.8061] | [0.7912] | [0.9160] |
| Controls | No | Yes | No | No | Yes | No | Yes | No | No | Yes |
| Method | OLS | OLS | OLS | OLS | OLS | MM | MM | MM | MM | MM |

Note: Table 3 focuses on identifying the determinants of Cumulative Abnormal Returns (CARs) during the full match period, employing both Ordinary Least Squares (OLS) and Robust MM-Estimator models. The table is structured into two main sections, each having several sub-models: "Returns (OLS)" and "Returns (Robust MM-Estimator)." In both main sections, the outcomes of various scenarios are presented—such as the effects of a (i) Victory, (ii) Victory under Low-Stake and (iii) High-Stake situations, (iv) Defeat, (v) Defeat under Low-Stake and (vi) High-Stake scenarios, and (vii) Defeat coupled with a Knockout. Each outcome is accompanied by its coefficient value and standard error in parentheses. For each model, we report two key fit statistics—R-squared ($R^2$) and Adjusted R-squared (Adj. $R^2$) for OLS models, while for Robust MM-Estimator models, pseudo R-squared ($Rw^2$) values are provided. The table includes controls in models (b), (e), (g) and (j) to account for other variables that may influence CARs. Statistical significance levels are represented with asterisks: * denotes $p < 0.10$, ** denotes $p < 0.05$, and *** denotes $p < 0.01$.

**Table 4. Determinants of Cumulative Abnormal Volumes During Full Match Period, OLS and Robust MM-Estimator Models**

| Period | (a) Coef. (SE) | (b) Coef. (SE) | (c) Coef. (SE) | (d) Coef. (SE) | (e) Coef. (SE) | (f) Coef. (SE) | (g) Coef. (SE) | (h) Coef. (SE) | (i) Coef. (SE) | (j) Coef. (SE) |
|---|---|---|---|---|---|---|---|---|---|---|
| **Outcome** | | | | | | | | | | |
| (i) Victory | 107.131*** (23.984) | 95.510*** (16.868) | | 107.131*** (24.679) | 83.064*** (12.339) | 85.642 (57.984) | 81.971** (36.852) | | 97.441 (70.268) | 81.204** (38.967) |
| (ii) Victory (Low-Stake) | | | -32.656 (94.006) | | | | | -33.599 (65.161) | | |
| (iii) Victory (High-Stake) | | | 0.242 (83.602) | | | | | -9.049 (67.995) | | |
| (iv) Defeat | 70.235* (35.517) | 78.822 (49.832) | | 100.602* (57.701) | 100.602 (65.983) | 45.424 (59.733) | 45.407 (37.963) | | 50.066 (75.704) | 47.577 (40.752) |
| (v) Defeat (Low-Stake) | | | -15.119 (111.020) | | | | | -68.226 (70.381) | | |
| (vi) Defeat (High-Stake) | | | -85.976 (101.648) | | | | | -78.988 (74.189) | | |
| (vii) Defeat & Knockout | | | | -70.857 (64.943) | -95.720 (94.915) | | | | -15.466 (51.716) | -10.126 (38.207) |
| $R^2$ | 0.128 | 0.418 | 0.143 | 0.204 | 0.491 | 0.142 | 0.466 | 0.079 | 0.175 | 0.475 |
| Adj. $R^2$ [$Rw^2$] | 0.031 | 0.168 | -0.071 | 0.063 | 0.216 | [0.317] | [0.849] | [0.228] | [0.301] | [0.846] |
| Controls | No | Yes | No | No | Yes | No | Yes | No | No | Yes |
| Method | OLS | OLS | OLS | OLS | OLS | MM | MM | MM | MM | MM |

Note: Table 4 focuses on identifying the determinants of Cumulative Abnormal Returns (CAVs) during the full match period, employing both Ordinary Least Squares (OLS) and Robust MM-Estimator models. The table is structured into two main sections, each having several sub-models: "Returns (OLS)" and "Returns (Robust MM-Estimator)." In both main sections, the outcomes of various scenarios are presented—such as the effects of a (i) Victory, (ii) Victory under Low-Stake and (iii) High-Stake situations, (iv) Defeat, (v) Defeat under Low-Stake and (vi) High-Stake scenarios, and (vii) Defeat coupled with a Knockout. Each outcome is accompanied by its coefficient value and standard error in parentheses. For each model, we report two key fit statistics—R-squared ($R^2$) and Adjusted R-squared (Adj. $R^2$) for OLS models, while for Robust MM-Estimator models, weighted R-squared ($Rw^2$) values are provided. The table includes controls in models (b), (e), (g) and (j) to account for other variables that may influence CAVs. Statistical significance levels are represented with asterisks: * denotes $p < 0.10$, ** denotes $p < 0.05$, and *** denotes $p < 0.01$.

$$CAR_{i,t} = \alpha_1 + \beta_1(Win_D) + \beta_2(Loss_D) + \sum_n \delta_n Z_{i,t} + \varepsilon_{i,t} \qquad (5)$$

To enhance our insights into the influence of match stakes on fan token CARS, we segment our dataset into two distinct categories: lower-stakes matches occurring during the group stage and higher-stakes football matches from the round of 16 to the finals.[14] Equation (6) introduces specific dummy variables – ($Low_D$) and ($High_D$) – that delineate these lower and higher stakes, respectively. This partitioning allows us to isolate the individual effects of victories and defeats on CARs while factoring in the contextual importance of the tournament's match stage.

$$CAR_{i,t} = \alpha_1 + Win_D[\beta_1(Low_D) + \beta_2(High_D)] \qquad (6)$$
$$+ Loss_D[\beta_3(Low_D) + \beta_4(High_D)] + \varepsilon_t$$

Estimates for Equation (6) are delineated in Table 3, employing both OLS (model c) and the robust MM-estimator (model h) methodologies. The coefficients, $\beta_1$ and $\beta_2$, are statistically insignificant, signifying that the fan token repercussions of victories are negligible, spanning both lower-stakes ($\beta_1$) and higher-stakes ($\beta_2$) matches. Conversely, outcomes stemming from defeats uncover compelling trends. A defeat in a lower-stakes match ($\beta_3$), precipitates a statistically insignificant impact ranging between -7.720% and -16.54% on fan token CARs. However, a defeat in a higher-stakes match ($\beta_4$) induces a highly significant plunge in fan token CARs, ranging from -39.35% to -57.70%, underscoring the pronounced sensitivity to defeats in matches of heightened importance.

The heightened impact of defeats in high-stakes matches compared to lower-stakes matches stems from several factors amplifying outcomes. Firstly, the pivotal do-or-die characteristic inherent to high-stakes matches intensifies their influence on fan token CARs, propelled by escalating emotional and temporal investments from both teams and fans. Interestingly, this emotional investment impacts CAVs differently, with victories exerting a more pivotal role in driving trading volumes than defeats. Victories in these critical matches generate excitement and positive sentiment that may stimulate trading activity, reflecting heightened investor engagement and optimism. Secondly, the enhanced media spotlight and broader public scrutiny during these decisive matches amplify the ramifications of the outcomes, elevating emotional stakes through heightened anticipations and pressures. The celebratory atmosphere following a victory and the desire to be part of the successful team's journey can catalyze investors to engage more actively in the market, thus elevating trading volumes. Thirdly, achievements during these critical junctions of the tournament often possess enduring impacts on a team's legacy, thus becoming pivotal milestones that inject added sensitivity into fan token metrics. In essence, while both victories and defeats in matches of paramount importance affect fan token metrics, our analysis uncovers a nuanced dynamic. Specifically, victories significantly influence CAVs, indicating a robust investor response to positive outcomes. However, defeats in high-stakes matches exert a more substantial impact on CARs, highlighting the intricate relationship between match outcomes and fan token market dynamics.

---

[14] This partitioning is informed by the distribution of matches, considering approximately half of all matches transpire in the lower stakes group stage while the remaining unfold in later stages; thus, we bifurcate the sample into these categories to achieve a more streamlined classification. A more granular segmentation would necessitate the incorporation of an excessive number of variables into the model.

To gain nuanced insights into the influence of high-stakes matches, specifically knockout matches, where defeats lead to the team's elimination from the World Cup, on fan token CARs, we refine our analytical framework, integrating two key Equations (7) and (8). Equation (7) introduces an interaction term between ($Loss_D$) and a dummy variable for knockouts ($Knockout_D$). To fortify the robustness of our analysis, Equation (8) further refines the model with control variables ($\sum_n \delta_n Z_{i,t}$), aligning with Equation (5). As all knockout matches in our dataset are inherently high-stakes matches, the model does not require explicit control to address impact in a low-stakes scenario; however, is captured by the β$_2$ coefficient. This meticulous methodology facilitates a comprehensive exploration of how pivotal junctures in the World Cup shape fan token CARs.

$$CAR_{i,t} = \alpha_1 + \beta_1(Win_D) + \beta_2(Loss_D) + \beta_3(Loss_D * Knockout_D) + \varepsilon_{i,t} \quad (7)$$

$$CAR_{i,t} = \alpha_1 + \beta_1(Win_D) + \beta_2(Loss_D) + \beta_3(Loss_D * Knockout_D) \\ + \sum_n \delta_n Z_{i,t} + \varepsilon_{i,t} \quad (8)$$

The influence of both victory (β$_1$) and general defeat (β$_2$) consistently proves to be statistically insignificant, robust to variations in model specifications (d), (e), (i), and (j), and aligns with earlier models outlined in Table 3, further emphasizing their negligible impacts on fan token CARs. In contrast, the interaction terms (β$_3$) emerge as statistically significant, displaying a negative coefficient across all models. This robust indication implies that defeats in knockout stages render a marked and deleterious effect on fan token CARs. More specifically, a defeat in a knockout scenario triggers a sharp additional decline in CARs by around -22.81% to -36.65% when analyzed using

OLS models (d) and (e). Most notably, this decline in CARs is more pronounced using robust MM-estimator models (i and j), showing a staggering additional -50.74% to 59.43% decline attributable to knockouts. These significant responses highlight the substantial ramifications that an intensified environment of knockout matches imposes on fan token valuation.

Examining CAVs, Table 4 reveals that victories ($\beta_1$) exert a robust, positive, and statistically significant influence on CAVs, robust to all model specifications. In contrast, defeats yield a more muted impact on CAVs. The influence of match stakes, whether low or high, emerges as statistically insignificant, indicating a lack of consequential effect on market reactions. The knockout variable shows negative yet statistically inconsequential responses throughout the models. The contrast between Table 3 and Table 4 stems from emotional and psychological commitment of fans and teams when assessing stock returns (CARs) against trading volumes (CAVs). In instances of high stakes, a defeat substantially depresses the value of fan tokens, evidenced by the CARs; however, does not necessarily inhibit trading activity, thus permitting CAVs to sustain less impact.

The increase in CAVs without a corresponding significant impact on CARs suggests that while victories energize the market, leading to heightened trading volumes, they do not necessarily translate into positive price movements. This discrepancy arises from complex market psychology dynamics. Firstly, the selling pressure from token holders realizing gains may neutralize the buying pressure from investors experiencing fear of missing out (FOMO), anticipating price increases. This balance of contrasting forces may explain the surge in trading activity without a proportional impact on token prices. Secondly, in the context of market efficiency, the rapid assimilation of match outcomes into fan token prices coupled with emotional and speculative trading following victories may explain the discrepancy as traders, driven more by sentiment and the desire to participate in the event's excitement and capitalize on the token's price volatility may supersede an informed assessment of the victory's impact on the token's intrinsic value.

## 5  Buy the Rumor, Sell the News

The counterintuitive findings revealing declining CARS over inter-match segments in Table 2 necessitate a more in-depth exploration of the foundational dynamics impacting fan token returns throughout World Cup matches. As theorized in Section 4.3, the unforeseen decline in fan token values from the commencement of each match may exemplify the longstanding market maxim "buy the rumor, sell the news" (Edmans et al., 2012). To empirically assess this proposition, we executed an event study, utilizing daily price data for the four national teams possessing fan tokens and participating in the World Cup, sourced from coingecko.com between January and December 2022. The outcomes, presented in Table 5, yield substantiation for a positive anticipation effect persisting throughout the three months leading up to the FIFA World Cup and uncovers significant and exceptionally elevated abnormal returns during and after the World Cup, consistent with the hypothesis.

In the six months before the World Cup (days -120 to -1), our event study revealed a surge in fan token returns, highlighted by a highly significant CAR of 211.26%. This performance implies that investors, drawn by the international focus on the World Cup, entered the market anticipating considerable profits. Nevertheless, this upbeat momentum began to wane as the World Cup neared. The intervals from days -60 to -1 and -30 to -1 exhibit diminishing, though still positive, CARs of 40.75% and 20.08%, respectively. Investors began optimistic; their enthusiasm moderated, potentially in anticipation of the uncertainty introduced by actual match outcomes.

Remarkably, from the initiation of the World Cup (days 0 to 26), CARs manifested a profound plunge, registering at -173.40% and continuing the steep decline in the ensuing period to the termination of the tournament (days 27 to 56), plummeting to -1,000.14%.[15] This pronounced downturn is congruent with our finding that token values deteriorate sharply from the commencement to the cessation of each match (Section 4.3). Further strengthening our hypothesis, the "Pos." column in Table 5 depicts 100% positive CARs preceding the World Cup, plunging to 0% amidst and following the event. This sharp transition in investor sentiment aligns with critical junctures in the World Cup timeline.

In summary, our findings underscore the impact of the "buy the rumor, sell the news" strategy on fan token market behavior during high-profile events like the World Cup. Investors capitalize on pre-event hype only to promptly liquidate positions as matches begin, likely as a risk-mitigation strategy. This pattern not only echoes trading dynamics seen in traditional financial markets but also enhances our understanding of the complex behavior of fan tokens during major sporting events.

---

[15] While, at a first glance, it might seem counterintuitive for CARs to exceed -100%, this is indeed accurate. It succinctly illustrates that, over a certain timeframe, the abnormal returns are substantially inferior to what was anticipated or expected. This phenomenon is a reflection of the profound deviations in actual returns from their expected values, denoting severe market contractions or substantial losses in the specified period.

**Table 5. Daily Fan Token Returns of National Teams around the World Cup**

| Days to the World Cup in t = 0 | CARs | Std. Err. | t-test | z-test | Pos. |
|---|---|---|---|---|---|
| -120 to -1 | 211.26% | 23.47% | 9.00*** | 2.92** | 100% |
| -60 to -1 | 40.75% | 22.48% | 1.81 | 1.46 | 75% |
| -30 to -1 | 20.08% | 16.33% | 1.23 | 0.73 | 50% |
| 0 to 26 (World Cup) | -173.40% | 37.82% | -4.59*** | -2.89** | 0% |
| 27 to 56 | -1,000.14% | 392.02% | -2.59* | -1.83* | 0% |

Note: *, **, *** indicate significance at the 10%, 5% and 1% level. N = 4. Cumulative abnormal returns (CARs) of the four tokens (ARG, BFT, POR, SNFT) are estimated on the basis of a market model with Bitcoin as reference market and an estimation period of 200 days. Pos. indicates the share of positive CARs. The z-test refers to the Wilcoxon sign rank test (Wilcoxon, 1945).

Since significant returns observed near the tournament suggest that market reactions may be more influenced by public attention than by fundamental changes, future studies could analyze market responses following teams' qualification outcomes. Similarly to the loss effect, the (unexpected) failure of teams to qualify for a major event like the World Cup could negatively impact fan token returns, while their (expected) qualification might not significantly affect the market. This disparity in market reaction could underscores the influence of expectations and team status, where negative surprises have a stronger impact than anticipated successes[16].

# 6   Conclusion and Future Directions

In the convergence of sports and financial landscapes, this paper explores the nuanced and multifaceted market reactions of fan token values and trading volumes to developments during 2022 World Cup matches. The paper reveals a stark contrast in market reactions to victories and defeats, wherein the latter induces a substantial, statistically significant negative impact, especially in high-stakes matches and knockout scenarios. This heightened sensitivity to defeats underscores the intense emotional and temporal commitments and the amplified ramifications due to increased public scrutiny and media spotlight. The influence of victories on fan token values remained statistically insignificant, highlighting the pronounced market predisposition towards losses.

Furthermore, the study offers evidence supporting the longstanding market maxim, "buy the rumor, sell the news." The fan token market experienced surges in the anticipation phase preceding the World Cup, followed by a profound plunge in value at the onset of each match. This pattern signifies investors' proclivity to capitalize on pre-event hype and subsequently liquidate positions as a risk-mitigation strategy as matches unfold, mirroring the trading dynamics prevalent in traditional financial markets.

---

[16] The specific analysis of the impact of failure to qualify (as observed with Italy and Peru) and the successful qualification on fan tokens (i.e., ARG, BFT, POR, SNFT) was constrained due to the recency of fan tokens. The necessary data basis for a detailed examination was not yet available, limiting the scope for a comprehensive analysis.

Our findings offer a captivating juxtaposition of behavioral finance theory and the efficient markets hypothesis (EMH) within the innovative domain of fan tokens. EMH theorizes the instantaneous incorporation of all public information into asset valuations (Fama, 1970), yet analysis reveals that the fan token markets are entwined with inefficiencies, markedly shaped by emotional biases. This aligns with the foundational tenets of behavioral finance, acknowledging the frequent deviations of investors from unadulterated rationality (Barberis and Thaler, 2003; Shleifer, 2000).

For practitioners, the insights gleaned from this study highlight the imperative of cognizance towards fan sentiment and match repercussions in trading fan tokens. Given the susceptibility of these markets to emotional fluctuations, a vigilant approach to their potential ramifications is essential, particularly amid high-stakes encounters. Moreover, sports entities issuing fan tokens must assimilate these dynamics in their communication strategies, especially concerning match results. Discerning the psychological impacts of football encounters on fan token valuations can empower organizations to navigate markets and fine-tune engagement paradigms.

References

Ajinkya, B.B., Jain, P.C., 1989. The behavior of daily stock market trading volume. J. Account. Econ. 11, 331–359. https://doi.org/10.1016/0165-4101(89)90018-9

Alexander, C., Dakos, M., 2020. A critical investigation of cryptocurrency data and analysis. Quant Finance 20, 173–188. https://doi.org/10.1080/14697688.2019.1641347

Ante, L., 2019. Market Reaction to Exchange Listings of Cryptocurrencies. https://doi.org/10.13140/RG.2.2.19924.76161

Ante, L., Schellinger, B., Wazinski, F.-P., 2023. Enhancing Trust, Efficiency, and Empowerment in Sports – Developing a Blockchain-based Fan Token Framework, in: Eu (Ed.), 2023 European Conference on Information Systems (ECIS 2023).

Ante, L., Schellinger, B., Demir, E., 2024a. The impact of football games and sporting performance on intra-day fan token returns. Journal of Business Economics. https://doi.org/10.1007/s11573-023-01187-z

Ante, L., Saggu, A., Schellinger, B., Wazinski, F.-P., 2024b. Voting Participation and Engagement in Blockchain-based Fan Tokens. Electronic Markets. Forthcoming.

Armitage, S., 1995. Event study methods and evidence on their performance. J. Econ. Surv. 9, 25–52. https://doi.org/10.1111/j.1467-6419.1995.tb00109.x

Ashton, J.K., Gerrard, B., Hudson, R., 2011. Do national soccer results really impact on the stock market? Appl. Econ. 43, 3709–3717. https://doi.org/10.1080/00036841003689762

Ashton, J.K., Gerrard, B., Hudson, R., 2003. Economic impact of national sporting success: evidence from the London stock exchange. Appl. Econ. Lett. 10, 783–785. https://doi.org/10.1080/1350485032000126712

Assaf, A., Demir, E., & Ersan, O. (2024). Detecting and date-stamping bubbles in fan tokens. International Review of Economics & Finance.

Barberis, N., Thaler, R., 2003. Chapter 18 A survey of behavioral finance. pp. 1053–1128. https://doi.org/10.1016/S1574-0102(03)01027-6

Bariviera, A.F., Merediz-Solà, I., 2021. Where do we stand in cryptocurrencies economic research? A survey based on hybrid analysis. J Econ Surv 35, 377–407. https://doi.org/10.1111/joes.12412

Bernile, G., Lyandres, E., 2011. Understanding Investor Sentiment: The Case of Soccer. Financ. Manag. 40, 357–380. https://doi.org/10.1111/j.1755-053X.2011.01145.x

Benedetti, H., & Kostovetsky, L. (2021). Digital tulips? Returns to investors in initial coin offerings. Journal of Corporate Finance, 66, 101786.

Boehmer, E., Musumeci, J., Poulsen, A.B., 1991. Event-study methodology under conditions of event-induced variance. J. Financ. Econom. 30, 253–272.

Curatola, G., Donadelli, M., Kizys, R., Riedel, M., 2016. Investor Sentiment and Sectoral Stock Returns: Evidence from World Cup Games. Financ. Res. Lett. 17, 267–274. https://doi.org/10.1016/j.frl.2016.03.023

Cong, L. W., Li, Y., & Wang, N. (2021). Tokenomics: Dynamic adoption and valuation. The Review of Financial Studies, 34(3), 1105-1155.

Cong, L. W., Li, Y., & Wang, N. (2022). Token-based platform finance. Journal of Financial Economics, 144(3), 972-991.

Demir, E., Danis, H., 2011. The Effect of Performance of Soccer Clubs on Their Stock Prices: Evidence from Turkey. Emerg. Mark. Financ. Trade 47, 58–70. https://doi.org/10.2753/REE1540-496X4705S404

Demir, E., Ersan, O., Popesko, B., 2022. Are Fan Tokens Fan Tokens? Financ. Res. Lett. 102736. https://doi.org/10.1016/j.frl.2022.102736

Demir, E., Rigoni, U., 2017. You Lose, I Feel Better: Rivalry Between Soccer Teams

and the Impact of Schadenfreude on Stock Market. J. Sports Econom. 18, 58–76. https://doi.org/10.1177/1527002514551801

Edmans, A., Garcia, D., Norli, Ø., 2007. Sports Sentiment and Stock Returns. J. Finance 62, 1967–1998. https://doi.org/10.1111/j.1540-6261.2007.01262.x

Ehrmann, M., Jansen, D.-J., 2016. It Hurts (Stock Prices) When Your Team is about to Lose a Soccer Match *. Rev. Financ. 20, 1215–1233. https://doi.org/10.1093/rof/rfv031

Ersan, O., Demir, E., Assaf, A., 2022. Connectedness among fan tokens and stocks of football clubs. Res. Int. Bus. Financ. 63, 101780. https://doi.org/10.1016/j.ribaf.2022.101780

Fama, E.F., 1970. Efficient Capital Markets: A Review of Theory and Empirical Work. J. Finance 25, 383–417. https://doi.org/10.2307/2325486

FIFA, 2022. Men's Ranking [WWW Document]. FIFA. URL https://www.fifa.com/fifa-world-ranking/men?dateId=id13792 (accessed 12.27.22).

Foglia, M., Maci, G., & Pacelli, V. (2024). FinTech and fan tokens: Understanding the risks spillover of digital asset investment. Research in International Business and Finance, 68, 102190.

Geyer-Klingeberg, J., Hang, M., Walter, M., Rathgeber, A., 2018. Do stock markets react to soccer games? A meta-regression analysis. Appl. Econ. 50, 2171–2189. https://doi.org/10.1080/00036846.2017.1392002

Howell, S. T., Niessner, M., & Yermack, D. (2020). Initial coin offerings: Financing growth with cryptocurrency token sales. The Review of Financial Studies, 33(9), 3925-3974.

Kaplanski, G., Levy, H., 2010. Exploitable Predictable Irrationality: The FIFA World Cup Effect on the U.S. Stock Market. J. Financ. Quant. Anal. 45, 535–553. https://doi.org/10.1017/S0022109010000153

Klein, C., Zwergel, B., Heiden, S., 2009a. On the existence of sports sentiment: the relation between football match results and stock index returns in Europe. Rev. Manag. Sci. 3, 191–208. https://doi.org/10.1007/s11846-009-0031-8

Klein, C., Zwergel, B., Henning Fock, J., 2009b. Reconsidering the impact of national soccer results on the FTSE 100. Appl. Econ. 41, 3287–3294. https://doi.org/10.1080/00036840802112471

Lopez-Gonzalez, H., Griffiths, M.D., 2023. Gambling-like Features in fan Tokens. J. Gambl. Stud. https://doi.org/10.1007/s10899-023-10215-0

MacKinlay, A.C., 1997. Event Studies in Economics and Finance. J. Econ. Lit. 35, 13–39.

Manoli, A.E., Dixon, K., Antonopoulos, G.A., 2024. Football Fan Tokens as a mode of "serious leisure": unveiling the dual essence of identity and investment. Leisure Studies 1–15. https://doi.org/10.1080/02614367.2024.2301949

Mazur, M., Vega, M., 2022. Football and Cryptocurrencies. https://doi.org/10.2139/ssrn.4035558

Meng, J., Fu, F., 2020. Understanding gambling behaviour and risk attitudes using cryptocurrency-based casino blockchain data. R. Soc. Open Sci. 7, 201446. https://doi.org/10.1098/rsos.201446

Nadler, P., & Guo, Y. (2020). The fair value of a token: How do markets price cryptocurrencies?. Research in International Business and Finance, 52, 101108.

Palomino, F., Renneboog, L., Zhang, C., 2009. Information salience, investor sentiment, and stock returns: The case of British soccer betting. J. Corp. Financ. 15, 368–387. https://doi.org/https://doi.org/10.1016/j.jcorpfin.2008.12.001

Renneboog, L.D.R., Vanraband, P., 2000. Share Price Reactions to Sporty Performances of Soccer Clubs listed on the London Stock Exchange and the AIM, CentER Discussion Paper.

Rocketfan, 2023. Fan Engagement Asset Market Cap [WWW Document]. URL https://rocketfan.com/market (accessed 5.5.23).

Saggu, A., 2022. The Intraday Bitcoin Response to Tether Minting and Burning Events: Asymmetry, Investor Sentiment, and "Whale Alerts" on Twitter. Financ. Res. Lett. 49, 103096. https://doi.org/10.1016/j.frl.2022.103096

Scharnowski, M., Scharnowski, S., Zimmermann, L., 2022. Fan Tokens: Sports and Speculation on the Blockchain. https://doi.org/10.2139/ssrn.3992430

Scholtens, B., Peenstra, W., 2009. Scoring on the stock exchange? The effect of football matches on stock market returns: an event study. Appl. Econ. 41, 3231–3237. https://doi.org/10.1080/00036840701721406

Shleifer, A., 2000. Inefficient Markets. Oxford University PressOxford. https://doi.org/10.1093/0198292279.001.0001

Truong, Q.-T., Tran, Q.-N., Bakry, W., Nguyen, D.N., Al-Mohamad, S., 2021. Football sentiment and stock market returns: Evidence from a frontier market. J. Behav. Exp. Financ. 30, 100472. https://doi.org/10.1016/j.jbef.2021.100472

Vidal-Tomás, D., 2023. Blockchain, sport and fan tokens. Journal of Economic Studies 51(1), 24-38. https://doi.org/10.1108/JES-02-2023-0094

Vidal-Tomás, D., 2022. Which cryptocurrency data sources should scholars use? International Review of Financial Analysis 81, 102061. https://doi.org/10.1016/j.irfa.2022.102061

Wann, D.L., Dolan, T.J., MeGeorge, K.K., Allison, J.A., 1994. Relationships

between Spectator Identification and Spectators' Perceptions of Influence, Spectators' Emotions, and Competition Outcome. J. Sport Exerc. Psychol. 16, 347–364. https://doi.org/10.1123/jsep.16.4.347

Wilcoxon, F., 1945. Individual Comparisons by Ranking Methods. Biometrics Bull. 1, 80–83.

Yousaf, I., Jareño, F., & Esparcia, C. (2022). Tail connectedness between lending/borrowing tokens and commercial bank stocks. International Review of Financial Analysis, 84, 102417.

Yousaf, I., Pham, L., & Goodell, J. W. (2023). The connectedness between meme tokens, meme stocks, and other asset classes: Evidence from a quantile connectedness approach. Journal of International Financial Markets, Institutions and Money, 82, 101694.

Yousaf, I., Assaf, A., & Demir, E. (2024). Relationship between real estate tokens and other asset classes: Evidence from quantile connectedness approach. Research in International Business and Finance, 69, 102238.

Yu, Y., Wang, X., 2015. World Cup 2014 in the Twitter World: A big data analysis of sentiments in U.S. sports fans' tweets. Comput. Human Behav. 48, 392–400. https://doi.org/10.1016/j.chb.2015.01.075